\documentclass{ws-p10x7a}

\usepackage{graphicx}

\def\fkm {P. Fisher, B. Kayser, and K.S. McFarland, in {\it Ann.\ Rev.\ Nucl.\ Part.\ Sci.} {\bf 49}, eds. C. Quigg, V. Luth, and P. Paul (Annual Reviews, Palo Alto, CA, 1999) p. 481.}
\def\phlb#1#2#3{{\it Phys.\ Lett.} {\bf B\,#1,} #2 (19#3)}
\def\plb2#1#2#3{{\it Phys.\ Lett.} {\bf B\,#1}, #2 (20#3)}
\def\prd2#1#2#3{{\it Phys.\ Rev.} {\bf D\,#1}, #2 (20#3)}
\def\prd#1#2#3{{\it Phys.\ Rev.} {\bf D\,#1}, #2 (19#3)}

\def\sss{\scriptscriptstyle}
\def\ra{\rightarrow}

\def\nmb{\overline{\nu_m}}

\def\ket#1{|\nu_#1\rangle}
\def\barp{{\hspace{-.5ex}\raise.35ex\hbox{${\sss (}$}}---{\raise.35ex\hbox{${\sss )}$}}}
\def\nubarp#1{\hbox{$\nu_#1$\kern-1.0em\raise1.4ex\hbox{\barp}}}
\def\dm2#1{\delta M^2_{\mbox{{\scriptsize #1}}}}
\newcommand{\fiup}{\phi(\nu_\mu \mbox{ Up})}
\newcommand{\fidn}{\phi(\nu_\mu \mbox{ Down})}
\newcommand{\Eq}[1]{Eq.~(\ref{eq#1})}
\newcommand{\beq}{\begin{equation}}
\newcommand{\eeq}{\end{equation}}

\begin{document}

\title{Neutrino Mass: The Present and the Future\thanks{}}

\author{Boris Kayser}

\address{National Science Foundation, 4201 Wilson Blvd., Arlington VA 22230, USA \\E-mail: bkayser@nsf.gov}

\twocolumn[\maketitle\abstract{
We argue that the evidence for neutrino mass is quite compelling. This mass
raises a number of questions, which we enumerate, about neutrinos. Then we focus on one of these questions---the issue of the possible neutrino mass spectra. In
particular, we explain that one can have a four-neutrino spectrum which does not
require significant sterile-neutrino involvement in either the atmospheric or
solar neutrino oscillations.}]

%\section{Guidelines}

\begin{figure}[htb]
\includegraphics[scale=0.6]{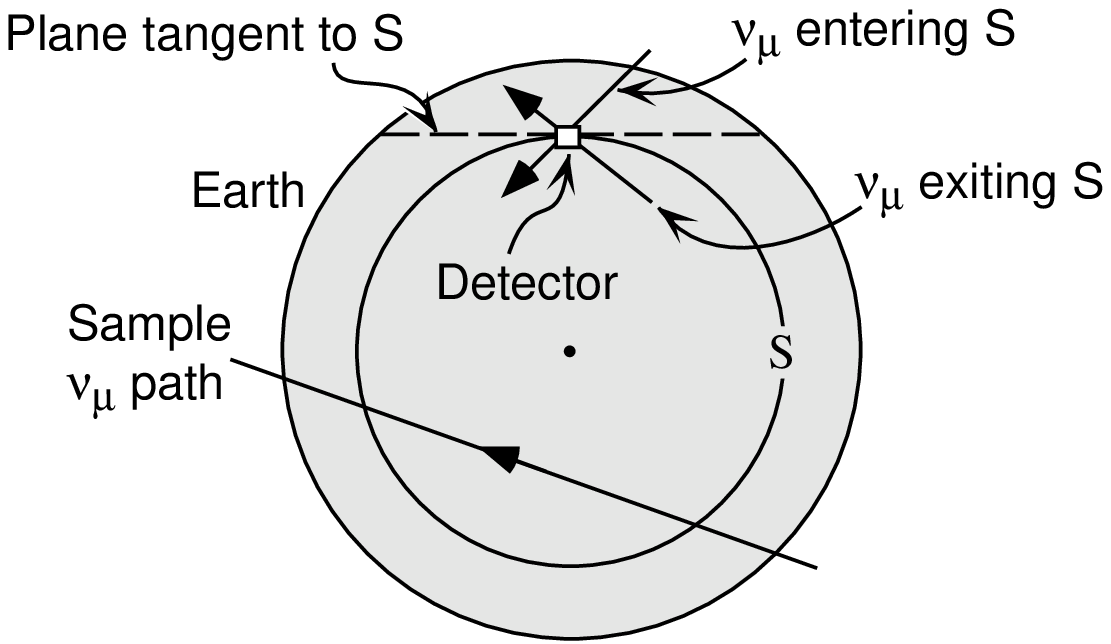}
\caption{Atmospheric muon neutrino fluxes at an underground detector. S is a sphere centered at the center of the earth and passing through the detector.}
\label{fA}
\end{figure}

\footnotetext{$^*$To appear in the Proceedings of the XXXth International Conference on High Energy Physics, held in Osaka, Japan, July 27--August 2, 2000.}
Before we discuss the physics of neutrinos with mass, let us step back and ask whether the evidence that neutrinos {\it do} have mass is really convincing. We believe that it is. The most compelling single piece of evidence is the observed violation of the equality
\beq
\fiup = \fidn \; .
\label{eq1}
\eeq
Here, $\fiup$ is the total flux of atmospheric muon neutrinos observed by an underground detector to be coming {\it upward} from all directions below the horizontal at the location of the detector, while $\fidn$ is the corresponding total flux observed to be coming {\it dowmward} from all directions above the horizontal. The atmospheric neutrinos are produced by cosmic rays in the earth's atmosphere all around the world, and so are incident on the detector from all directions. 
In considering our expectations for the relationship between $\fiup$ and $\fidn$, let us suppose that nothing---neither neutrino oscillation nor anything else---decreases or increases the atmospheric $\nu_\mu$ flux as the neutrinos travel from their points of origin to the detector. Then, as illustrated by the ``Sample $\nu_\mu$ path'' in Fig.~\ref{fA}, any $\nu_\mu$ that enters the sphere $S$ defined in the figure caption will eventually exit this sphere. 
Thus, since we are dealing with a steady-state situation, the total $\nu_\mu$ fluxes entering and exiting $S$ per unit time must be equal. Now, for neutrino energies $E >$ a few GeV, the flux of cosmic rays that create the atmospheric neutrinos is known to be isotropic. Thus, at these energies, the atmospheric muon neutrinos are being produced at the same rate everywhere around the earth. Thanks to this spherical symmetry, the equality betwen the $\nu_\mu$ fluxes entering and exiting $S$ must hold, not only for $S$ as a whole, but at each point of $S$. 
In particular, it must hold at the location of the detector. But, as is clear from Fig.~\ref{fA}, a $\nu_\mu$ entering $S$ through the detector must be part of the downward flux $\fidn$. One exiting $S$ through the detector must be part of $\fiup$. Thus, the equality of the $\nu_\mu$ fluxes entering and exiting $S$ at the detector implies that $\fidn = \fiup$.\cite{r1} With a bit more effort, but no additional assumptions, one can show that this equality must hold not only for the integrated downward and upward fluxes, but angle by angle. That is, the flux coming down from zenith angle $\theta_z$ must equal that coming up from angle $\pi - \theta_z$.\cite{r2}

The Super-Kamiokande detector (Super-K) finds that for multi-GeV atmospheric muon neutrinos,\cite{r3}
\begin{eqnarray}
\frac{\phi (\nu_\mu \mbox{ Up}; -1.0<\cos\theta_z < -0.2)}{\phi (\nu_\mu \mbox{ Down}; +0.2<\cos\theta_z < +1.0)} &  \nonumber \\
 & \hspace{-1.5in} = 0.54 \pm 0.04 \; ,
\label{eq2}
\end{eqnarray}
in strong disagreement with the expected equality of upward and downward fluxes. Thus, some mechanism must be changing the atmospheric $\nu_\mu$ flux while the neutrinos travel to the detector. As we see, this conclusion follows merely from the isotropy of the cosmic rays, the fact that the earth is round, and the fact that the ratio in \Eq{2} is not unity.

The most attractive candidate for the mechanism that is altering the atmospheric $\nu_\mu$ flux is the oscillation of the muon neutrinos into neutrinos of another flavor. Indeed, neutrino oscillation fits the detailed atmospheric neutrino data very well.\cite{r3} Barring the exotic (albeit intriguing) possibility of extra spatial dimensions, neutrino oscillation implies neutrino mass.

Amusingly, an alternative candidate, neutrino decay, also fits the detailed atmospheric neutrino data well.\cite{r4} To be sure, decay within the time that a neutrino takes to traverse the earth is theoretically less likely than oscillation. Nevertheless, it is interesting that the decay model\cite{r4} survives all the comparisons with data that have so far been made. Future long-baseline neutrino experiments capable of distinguishing between the sinusoidal dependence on (distance/energy) that is characteristic of oscillation and the exponential dependence that is characteristic of decay would discriminate between the two possibilities. 
The decay hypothesis would also be tested by more accurate information on the rate of neutral current (NC) events induced by atmospheric neutrinos in an underground detector. If $\nu_\mu$ oscillates to $\nu_\tau$, the NC event rate will be the same as if there were no oscillation or decay. But if neutrino decay is playing a prominent role, then the electroweak-active neutrino flux is reduced by the decay process, and so the NC event rate will be lower than when there is no oscillation or decay.\cite{r5} 

Both the oscillation and decay explanations of the behavior of atmospheric neutrinos imply neutrino mass and mixing. Strong further evidence for mass and mixing comes from the behavior of the solar neutrinos, which can be successfully explained in terms of matter-enhanced or perhaps vacuum oscillation.\cite{r3} Finally, there is unconfirmed evidence for $\nubarp{\mu} \ra \nubarp{e}$ oscillation in the LSND experiment.\cite{r6} As we have seen, the evidence for mass and mixing from the atmospheric neutrinos is very strong indeed.

That neutrinos have mass means that there is some spectrum of three or more neutrino mass eigenstates, $\nu_1, \nu_2, \nu_3, \ldots$, which are the neutrino analogues of the charged-lepton mass eigenstates, $e, \mu$, and $\tau$. That neutrinos mix means that the neutrino state $\ket{\ell}$ coupled by the weak interaction to the particular charged-lepton mass eigenstate $\ell$ (e, $\mu$, or $\tau$) is not one of the neutrino mass eigenstates $\ket{m}$, but some linear combination of the neutrino mass eigenstates. That is,
\beq
\ket{\ell} = \sum_m U^*_{\ell m} \ket{m} \; ,
\label{eq3}
\eeq
where $U$ is the unitary leptonic mixing matrix, often called the Maki-Nakagawa-Sakata matrix.\cite{r7}. The neutrino $\ket{\ell}$ is called the neutrino of ``flavor'' $\ell$. The decays $Z \ra \nu_\ell \overline{\nu_\ell}$ are known to produce only three distinct neutrinos of definite flavor: $\nu_e, \nu_\mu$, and $\nu_\tau$. However, there may be more than three neutrinos $\nu_m$ of definite {\em mass}. If, for example, there are four neutrino mass eigenstates, then one linear combination of them,
\beq
\ket{S} = \sum_m U^*_{sm}\ket{m} \; ,
\label{eq4}
\eeq
must not couple to the $Z$, and hence must not enjoy normal weak interactions. Consequently, this linear combination is referred to as a ``sterile'' neutrino.

Having learned that neutrinos almost certainly have mass and mix, we would like to learn the answers to the following questions:
\begin{itemize}
\item How many neutrino flavors, active and sterile, are there? Equivalently, how many neutrino mass eigenstates are there?
\item What are the masses, $M_m$, of the mass eigenstates $\nu_m$?
\item Is each neutrino of definite mass a Majorana particle ($\nmb = \nu_m$), or a Dirac particle  ($\nmb \neq \nu_m$)?
\item What are the elements $U_{\ell m}$ of the leptonic mixing matrix?
\item Does the behavior of neutrinos, in oscillation and other contexts, violate CP invariance?
\item What are the electromagnetic properties of neutrinos? In particular, what are their dipole moments?
\item What are the lifetimes of the neutrinos?
\end{itemize}

What we already know about these questions, and how we might learn more, are discussed in a previous paper.\cite{r8} Here, we would only like to add to that discussion some comments on the possible neutrino mass spectra and mixings suggested by the data on oscillation.

It is generally believed that if the atmospheric, solar, and LSND neutrinos all genuinely oscillate, then nature must contain at least four nondegenerate neutrino mass eigenstates.\cite{r9} As explained previously, the four corresponding neutrino flavor eigenstates must then be $\nu_e, \nu_\mu, \nu_\tau$, and a neutrino which is sterile, $\nu_S$. Thus, if the atmospheric, solar, and LSND oscillations are all genuine, then nature contains a fourth neutrino quite different from the three neutrinos already familiar to us.

\begin{figure}[htb]
\includegraphics[scale=0.5]{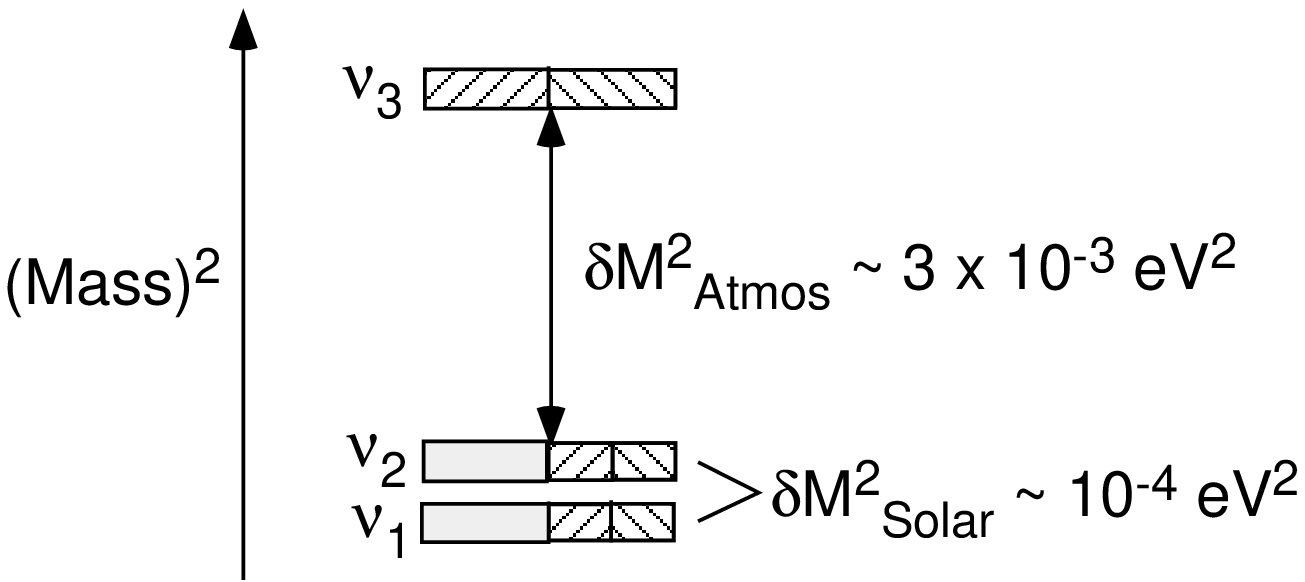}
\caption{A three-neutrino (Mass)$^2$ spectrum that accounts for the atmospheric and solar neutrino oscillations. The neutrinos $\nu_1, \nu_2$, and $\nu_3$ are mass eigenstates. The rough flavor content of each is indicated as follows: The $\nu_e$ fraction of a mass eigenstate is dotted, the $\nu_\mu$ fraction is shown by right-leaning hatching, and the $\nu_\tau$ fraction by left-leaning hatching.}
\label{f1}
\end{figure}

If the so-far unconfirmed oscillation seen in the LSND experiment is set aside, then the oscillations of the atmospheric and solar neutrinos can be explained in terms of just three neutrinos. The (Mass)$^2$ spectrum of these neutrinos can, for example, be as shown in Fig.~\ref{f1}. 
The height of this entire spectrum above (Mass)$^2 = 0$ is completely undetermined, because neutrino oscillation probabilities depend only on  (Mass)$^2$ {\em splittings} and not on the individual underlying masses.\cite{r8} The splitting $\dm2{Atmos} \sim 3 \times 10^{-3}$ eV$^2$ between mass eigenstates $\nu_3$ and $\nu_2$ is chosen to yield the observed atmospheric neutrino oscillation. The smaller splitting $\dm2{Solar} \sim 10^{-4}$ eV$^2$ between $\nu_2$ and $\nu_1$ is chosen, in this example, to be consistent with large-mixing-angle MSW neutrino flavor conversion in the sun. The flavor content of the mass eigenstates is chosen in the same way.\cite{r10} An alternative spectrum in which the two closely-spaced mass eigenstates are at the top of the picture, rather than at the bottom, is also possible.

\begin{figure}[htb]
\includegraphics[scale=0.55]{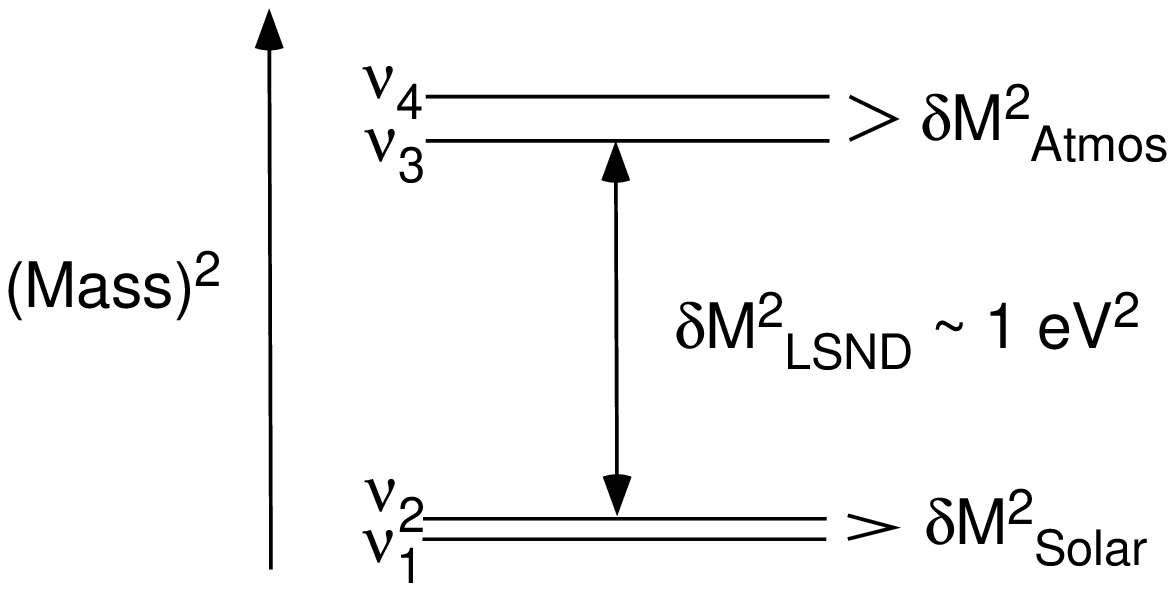}
\caption{A four-neutrino spectrum of the ``2+2'' variety. The neutrinos $\nu_1, \nu_2, \nu_3$, and $\nu_4$ are mass eigenstates. The splitting $\dm2{LSND}$ is the one called for by the LSND oscillation. An alternate spectrum with $\dm2{Solar}$ at the top and $\dm2{Atmos}$ at the bottom is also possible.}
\label{f2}
\end{figure}

\begin{figure}[htb]
\includegraphics[scale=0.6]{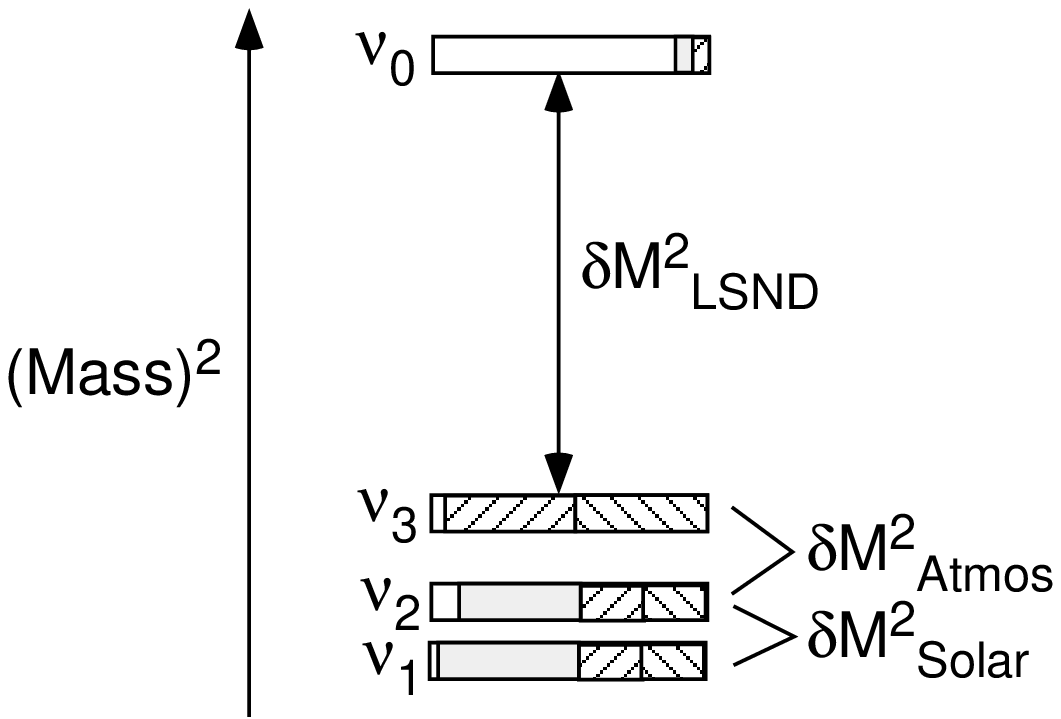}
\caption{A ``3+1'' spectrum consistent with the neutrino oscillation data. The neutrinos $\nu_1, \nu_2, \nu_3$, and $\nu_0$ are mass eigenstates. Their active flavor content is indicated as in Fig.~\ref{f1}, and their sterile flavor content by white regions. The very small active content of $\nu_0$, and the very small sterile content of $\nu_1$ -- $\nu_3$, are exaggerated.}
\label{f3}
\end{figure}

If we try to explain all reported oscillations, including the one seen by LSND, then, as already stated, the neutrino spectrum must contain at least four states. Until recently, it has been argued that, to be consistent with all oscillation data, both positive and negative, any such four-neutrino spectrum must be of the ``2+2'' variety.\cite{r11} That is, as illustrated in Fig.~\ref{f2}, it must consist of two pairs of neutrinos, with the members of each pair closely spaced, and with an ``LSND gap'' of order 1 eV$^2$ between the two pairs.
As previously explained, whenever there are four neutrino mass eigenstates, one linear combination of them must be a sterile neutrino, $\nu_S$. An interesting feature of the ``2+2'' four-neutrino schemes is that they predict that $\nu_S$ plays a significant role either in the atmospheric neutrino oscillation or in the solar one. However, analyses of the Super-K atmospheric neutrino data disfavor atmospheric neutrino oscillation into a sterile neutrino at the 99\% confidence level,\cite{r12} and are fully compatible with oscillation into an active neutrino. Furthermore, recent Super-K analyses of all the solar neutrino data disfavor solar neutrino oscillation into a sterile neutrino (either by the MSW effect or in vacuum).\cite{r3} 
Thus, at least to some degree, the data disfavor a major involvement of $\nu_S$ in either the atmospheric or solar oscillation.\cite{r13} This raises an interesting question: Suppose that, indeed, neither the neutrino state into which the atmospheric neutrinos oscillate, nor the one into which the solar ones do, is to any significant extent sterile. Would that rule out {\em all} four-neutrino explanations of the neutrino oscillation data? The answer to this question is ``no''.\cite{r14} The LSND experiment is now reporting\cite{r6} a somewhat lower oscillation probability than it did earlier. Thanks to this lower value, it is now possible to account for all the oscillation data with the ``3+1'' four-neutrino spectum shown in Fig.~\ref{f3}.\cite{r14}
This spectrum contains three light, relatively closely-spaced mass eigenstates, $\nu_1, \nu_2$, and $\nu_3$. These mass eigenstates are essentially fully active, and explain the atmospheric and solar oscillations in the same way as the three neutrinos in Fig.~\ref{f1} do. Thus, no sterile neutrino plays a significant role in either of these oscillations. However, the spectrum of Fig.~\ref{f3} also contains a fourth mass eigenstate, $\nu_0$, which is almost totally sterile, and which has a (Mass)$^2$ separated from those of $\nu_1, \nu_2$, and $\nu_3$ by an LSND gap of order 1 eV$^2$. In the past, a 3+1 spectrum of this kind was excluded by an incompatibility between LSND and the negative searches for $\nu_e$ and $\nu_\mu$ disappearance. 
To understand this incompatibility, we note that if the spectrum is like that of Fig.~\ref{f3}, then the mass splittings between $\nu_1, \nu_2$, and $\nu_3$ are invisible in any oscillation experiment with a distance to energy ratio, $L/E$, like that of LSND, for which $\dm2{Atmos, Solar} \times (L/E) << 1$. Thus, in any such experiment, there seem to be only two neutrinos: $\nu_0$, and the $\nu_1$-$\nu_2$-$\nu_3$ complex, whch appears to be only one neutrino. Hence, for any such experiment, the probabilities $P\,(\nu_\ell \ra \nu_{\ell ^{\sss\prime}})$ of the oscillations $\nu_\ell \ra \nu_{\ell ^{\sss\prime}}$ are described by the two-neutrino formulae\cite{r15}
\begin{eqnarray}
\lefteqn{P\,(\nu_\ell \ra \nu_{\ell ^{\sss\prime}\neq \ell}) = 
    4 P_\ell P_{\ell ^{\sss\prime}}} \nonumber \\
 & &    \times \sin^2 \left[1.27 \delta M^2\,(\mbox{eV}^2) \frac{L\,\mbox{(km)}}{E\,\mbox{(GeV)}}\right] 
\label{eq5}
\end{eqnarray}
and
\begin{eqnarray}
\lefteqn{P\,(\nu_\ell \ra \nu_\ell) = 1 - 4 P_\ell (1 - P_\ell)} \nonumber \\
 & &   \times \sin^2 \left[1.27 \delta M^2\,(\mbox{eV}^2) \frac{L\,\mbox{(km)}}{E\,\mbox{(GeV)}}\right] \; .
\label{eq6}
\end{eqnarray}
Here, $P_\ell \equiv |U_{\ell H}|^2$, where $\nu_H$ is the heavier of the two neutrino mass eigenstates, and $\delta M^2$ is the (Mass)$^2$ splitting between these eigenstates. For the ``quasi-two-neutrino''spectrum of Fig.~\ref{f3}, $\delta M^2= \dm2{LSND}$ and $P_\ell = |U_{\ell 0}|^2$. In particular, $P_e$ is the $\nu_e$ (dotted) fraction of $\nu_0$, and $P_\mu$ is the $\nu_\mu$ (right-leaning hatched) fraction. From \Eq{5}, we see that for any assumed value of $\dm2{LSND}$, the LSND $\nubarp{\mu} \ra \nubarp{e}$ oscillation determines an allowed range for $P_e P_\mu$. But, from \Eq{6}, we see that for the same assumed (Mass)$^2$ splitting, the negative searches at reactors for $\nu_e$ disappearance through oscillation place an upper limit on $P_e$. Similarly, the negative searches at accelerators for $\nu_\mu$ disappearance place an upper limit on $P_\mu$. 
With the $\nubarp{\mu} \ra \nubarp{e}$ oscillation probability reported earlier by LSND, there was no value of $\dm2{LSND}$ for which the LSND-allowed range for $P_e P_\mu$ was not incompatible with the upper limit on $P_e P_\mu$ coming from the negative searches for $\nu_e$ and $\nu_\mu$ disappearance. However, with the new, smaller oscillation probability being reported now by LSND, there are several values of $\dm2{LSND}$ for which the LSND-allowed range for $P_e P_\mu$ and the upper limit on this quantity from the negative searches for disappearance are not incompatible. Thus, the 3+1 spectrum of Fig.~\ref{f3} is a possible explanation of all the present neutrino oscillation data, even though it does not imply substantial sterile-neutrino involvement in either the atmospheric or solar neutrino oscillations. It will be interesting to see whether this spectrum can withstand future tests.

In conclusion, the evidence for neutrino mass has become quite convincing. However, we are just beginning to learn how many neutrinos there are, whether there are any sterile neutrinos, and what the neutrino masses and mixings are. While oscillation data already constrain the neutrino mass spectrum and neutrino mixing, a fair number of possibilities remain. In neutrino physics, interesting years lie ahead.

\section*{Acknowledgments}
It is a pleasure to thank the Aspen Center for Physics for its hospitality while some of the work reported here was done.

\end{document}